# Bell-inequality violation with a triggered photon-pair source

R. J. Young[1], R. M. Stevenson[1], A. J. Hudson[1,2], C. A. Nicoll[2], D. A. Ritchie[2], A. J. Shields[1]

[1]Toshiba Research Europe Limited, 208 Cambridge Science Park, Cambridge CB4 0GZ, United Kingdom

[2]Cavendish Laboratory, University of Cambridge, JJ Thompson Ave., Cambridge CB3 0HE, United Kingdom

**Abstract**

Here we demonstrate, for the first time, violation of Bell's inequality using a triggered quantum dot photon-pair source without post-selection. Furthermore, the fidelity to the expected Bell state can be increased above 90% using temporal gating to reject photons emitted at times when collection of uncorrelated light is more probable. A direct measurement of a CHSH Bell inequality is made showing a clear violation, highlighting that a quantum dot entangled photon source is suitable for communication exploiting non-local quantum correlations.

Quantum entanglement is an intriguing quirk of quantum theory linking the properties of spatially separated objects. Applied to pairs of photons transmitted through optical fibre or free space, it may be used to implement quantum communication applications such as quantum key distribution, quantum teleportation or entanglement swapping (for a recent review see [1]). Bell's inequality was devised to show that information stored locally in the entangled system does not dictate the outcome of measurements [2]. It is a more rigorous test for entanglement than the presence of non-classical correlations, requiring a higher fidelity to a maximally entangled state, and is an essential threshold for many applications exploiting spatially separated entangled photons, for example in ensuring the security of quantum communications [3]. Here, a triggered quantum dot source of entangled photons suitable for use in applications requiring high fidelities, such as those which exploit non-local correlations, is demonstrated by violating Bell's inequality.

A semiconductor quantum dot is an attractive source of single entangled photon pairs, [4-6] facilitating integration into optoelectronic devices. A pair of entangled photons is emitted from the quantum dot after first exciting two electrons and two holes into the dot with a short laser pulse to form a biexciton (*XX*). This state is spin neutral and has two optical decay paths to the ground state through intermediate exciton states (*X*) with a total angular momentum of one and opposing orientations. The wavefunction describing the emitted two-photon state is inseparable and hence the photons are entangled. The expected maximally entangled state can be expressed as $\psi^+=(|L_1R_2\rangle+|R_1L_2\rangle)/\sqrt{2}$ where the subscripts 1 and 2 indicate the first and second photons emitted, and *R (L)* refers to the right- (left-) circular polarisation of the emitted photons [Fig. 1].

This maximally-entangled state is only achieved if there is no 'which-path' information present in the decay process [7-9]. Typically an energetic splitting (*S*) between the two intermediate exciton states provides such information, though there has been a significant effort made by a number of groups worldwide to reduce the exciton level splitting. These include the simple selection of dots with small splitting [6, 10, 11], rapid thermal annealing of samples containing quantum dots [10, 12, 13], applying an in-plane electric [14, 15], magnetic [16, 17] or strain [18] field, and post-emission spectral filtering of the photons [5].

The sample used in this experiment consists of a self-organised low density layer of small InAs dots grown by molecular beam epitaxy in a GaAs matrix. The dot layer lies in a planar cavity between two Bragg mirrors consisting of alternating layers of $Al_{0.98}Ga_{0.02}As$ and GaAs. The cavity enhances the proportion of light emitted by a single dot, which is collected with a microscope objective (numerical aperture of 0.5) placed directly above the sample, by a factor of >10. The sample was studied at 10K in a continuous-flow helium-cooled cryostat with a silica window providing optical access. Electrons and holes were excited in the sample using a pulsed diode laser with a repetition rate of 80MHz, a pulse width of ~100ps and a photon energy greater than the GaAs bandgap energy.

The photoluminescence spectrum from a single quantum dot selected with a small splitting between its intermediate exciton states is shown in Fig. 2. Emission from the neutral exciton and biexciton states are labelled.

Two-photon emission from the biexciton cascade with a sufficiently high fidelity to violate a Bell inequality, without post-selection of the photons, requires a fine structure splitting of <0.5µeV [19]. We have developed a novel technique to measure exciton splittings on this scale despite an inhomegeous broadening of the excitons' emission which is around two orders of magnitude larger than such splittings. A half-waveplate was rotated in the path of the emission followed by a linear polarizer. For each waveplate angle the difference in energy between the exciton and biexciton photons ($E_{X-XX}$) was measured. The inset to Fig. 2 shows the result of this measurement relative to the angle average energy ($\overline{E_{X-XX}}$) to correct for systemic errors associated with beam displacement and sample movements. The difference in energy between the exciton and biexciton photons is seen to oscillate as a function of the waveplate angle with an amplitude of twice the exciton splitting (*2S*). The dot selected for this experiment shown in Fig. 2 has an exciton splitting S=0.32±0.06µeV which exceeds the requirement to test a Bell inequality without post-selection.

The degree of polarisation correlation between the biexciton and exciton photons is defined by

$$C_\mu = \frac{g_{2,1}^{(2)}(0) - g_{2,\bar{1}}^{(2)}(0)}{g_{2,1}^{(2)}(0) + g_{2,\bar{1}}^{(2)}(0)}$$

Where $g_{2,1}^{(2)}(0)$ is the zero-delay second-order correlation between the first and second photons in the polarisation basis $\mu$. $g_{2,\bar{1}}^{(2)}(0)$ is measured simultaneously with $g_{2,1}^{(2)}(0)$ and is the cross-polarised second order correlation. The dot emission was verified to be unpolarised with an error of 1.8%.

The fidelity of the state emitted by the quantum dot with respect to the expected state is calculated by measuring three parameters [19]:

$$f(\psi^+)=(1+C_{rectilinear}+C_{diagonal}-C_{circular})/4$$

Here $C_{rectilinear}$, $C_{diagonal}$ and $C_{circular}$ are the degrees of polarisation correlation between the pair of photons in the rectilinear, diagonal and circular bases respectively. $C_{rectilinear}$ is measured with linear polarisers in the measurement system, a half-wave plate is placed in the path of the dot emission at 22.5° to the lab-vertical for $C_{diagonal}$ and this is replaced with a quarter-wave plate at 45° to the lab-vertical for $C_{circular}$.

The zero-delay values of the second-order correlations used to calculate $C_{rectilinear}$, $C_{diagonal}$ and $C_{circular}$ are shown in blue in Fig. 3. These give $f(\psi^+)$=79.4±1.0%. This raw measured fidelity to the expected state is well in excess of the limit possible with any classical source, as well as the value required to test non-locality, a first for an entangled photon source of this type.

Without testing the non-locality of quantum mechanics the Bell parameter can be calculated with just two measurements for an unpolarised source, derived as follows. The probability that a first and second photon are detected with polarizations $\alpha$ and $\beta$ is given by $P(\alpha,\beta)=\langle\psi_{\alpha,\beta}|\rho|\psi_{\alpha,\beta}\rangle$, where $\psi_{\alpha,\beta}$ is the detected two-photon state vector, and $\rho$ the two-photon density matrix. Four such probabilities incorporating the orthogonal polarisations $\bar{\alpha}$ and $\bar{\beta}$ define the correlation function:

$$E(\alpha,\beta)=P(\alpha,\beta)+P(\bar{\alpha},\bar{\beta})-P(\alpha,\bar{\beta})-P(\bar{\alpha},\beta)$$

The Bell parameter expressed in the CHSH form [20] is:

$$S=E(\alpha,\beta)-E(\alpha',\beta)+E(\alpha,\beta')+E(\alpha',\beta')\leq 2$$

Maximal violation of the inequality is expected for combinations of circular and diagonal bases for α, and elliptical bases for β. The bases are equivalent to the well known optimum linear bases in conjunction with a quarter wave plate. *S* is readily expressed in terms of the matrix elements of ρ by substitution. Given generic ρ that is Hermitian with unity trace it is straightforward to show that $S = \sqrt{2}(E(D,D) - E(C,C))$, where the correlation functions in the diagonal and circular bases *E*(*D,D*) and *E*(C,*C*) are equivalent to the degrees of correlation $C_{diagonal}$ and $C_{circular}$ for an unpolarized source.

Three different Bell parameters corresponding to measurements in orthogonal planes of the Poincaré sphere are therefore given by:

$$S_{RC} = \sqrt{2}(C_{rectilinear} - C_{circular}) \leq 2$$
$$S_{DC} = \sqrt{2}(C_{diagonal} - C_{circular}) \leq 2$$
$$S_{RD} = \sqrt{2}(C_{rectilinear} + C_{diagonal}) \leq 2$$

The last of these expressions appeared in [21] during the preparation of this manuscript.

For the emission from the quantum dot we calculate *$S_{RC}$=2.15±0.05, $S_{DC}$=2.12±0.05, $S_{RD}$=1.88±0.04*. Two of these parameters show Bell violations of 2-3 standard deviations using the raw measured data. For this particular dot $C_{circular}$>$C_{rectilinear}$, $C_{diagonal}$; this is most likely a result of a weak coupling between the intermediate exciton states. The basis in which *C* is largest is found to vary for different quantum dots.

The degree to which Bell's inequality is violated is mainly limited by a small amount of uncorrelated light. A number of mechanisms are responsible for this. A fraction of emission originating from the InAs wetting layer is collected with the dot emission. The re-excitation probability is also non-zero due to the significance of the 100ps excitation pulse width relative to the 400ps lifetime of the biexciton state. Finally the dark count rate of the silicon avalanche photodiodes is inevitably non-zero.

The potential of the quantum dot source can be explored by applying a temporal gate to the emission. Photons arriving outside of a 1ns window centred on the biexciton emission and a 1.5ns window centred on the exciton emission are now discarded. Exclusion of photons emitted during or shortly after the excitation pulse reduces the contribution from multiple excitation of the dot and background emission that tends to have a shorter lifetime. As the dark count rate is time-independent, the signal to noise ratio is diminished for photons which arrive late in the emission period is diminished.

Using these temporal gates, the degree of polarisation correlation measured in the bases used earlier increases; the results are shown in red in Fig. 3. We now re-measure the three Bell parameters with the temporal gate: $S_{RC}$=2.51±0.12, $S_{DC}$=2.53±0.12, $S_{RD}$=2.44±0.09. Now all three parameters show violations of over 4 standard deviations. The fidelity with $(|L_1R_2>+|R_1L_2>)/\sqrt{2}$ is increased to 91.2±2.4%.

A four-parameter CHSH [20] measurement of Bell's inequality is achieved by placing independent half-waveplates in the detection channels for the exciton and biexciton photons. This allows non-orthogonal correlation measurements between the exciton and biexciton photons to be made. Waveplate angles with respect to the lab-vertical of 0°, 22.5° in the exciton detection channel and 11.25°, 33.75° in the biexciton channel were chosen and correlation measurement made in all four permutations of these angles. The results are presented in Table 1. The measured value of $C_{CHSH}$ of 2.45±0.10 agrees well with the two-parameter measurement $S_{RD}$=2.44±0.09.

The gating technique used here could be incorporated into applications though it is unlikely to be necessary given the high raw quality of the source. Even higher fidelities could be achieved by focusing on the degrading mechanisms highlighted by our gating technique. At early emission times the fidelity measured is limited by relatively slow population of the dot. A laser with shorter pulses, resonant electrical injection or quasi-resonant optical excitation could be used to remove the need for rejecting these photons with early arrival times. A Purcell enhancement of the radiative lifetime [22] would narrow the temporal window in which the majority of the photons are emitted from the quantum dot, which would exponentially decrease the number of photons rejected with late arrival times.

In summary, we have demonstrated triggered two-photon emission from the biexciton decay of a quantum dot and shown the two photons to be entangled in the Bell state $\psi^+=(|L_1R_2>+|R_1L_2>)/\sqrt{2}$ with a fidelity >90%. We have introduced three new measurements of Bell's parameter, showing that the raw emission of our device is of sufficient quality to be suitable for applications exploiting non-local quantum correlations, such as quantum key distribution. Finally we have performed a four-parameter measurement of Bell's inequality, which is violated by 4.5 standard deviations.

The authors would like to thank the EPSRC QIP IRC, EC FP6 Network of Excellence SANDiE and QAP for financial support.

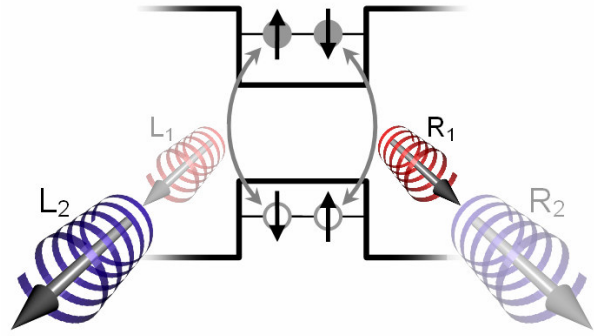

Figure 1. The two-photon biexciton decay process of a quantum dot with no bright exciton splitting. Recombination of a spin-up (-down) electron and a spin-down (-up) hole generates a left (right) hand circularly polarised photon. Since it is impossible to predict which photon will be emitted first, the emission is expected in the entangled state $(|L_1R_2\rangle+|R_1L_2\rangle)/\sqrt{2}$. Subscripts denote the order of emission of the left, L and right, R polarised photons.

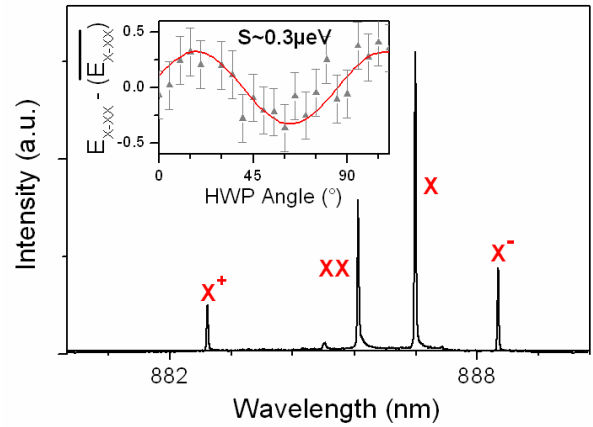

Figure 2. A single quantum-dot photoluminescence spectrum. Emission lines corresponding to electron-hole recombination from the neutral exciton (X), biexciton (XX), positively-charged exciton (X+) and negatively charged exciton (X-) are labelled. The energetic difference between the X and XX emission lines ($E_{X-XX}$) offset by its average is plotted as a function of half-waveplate angle to reveal a bright structure splitting (S) for this dot of ~0.3μeV.

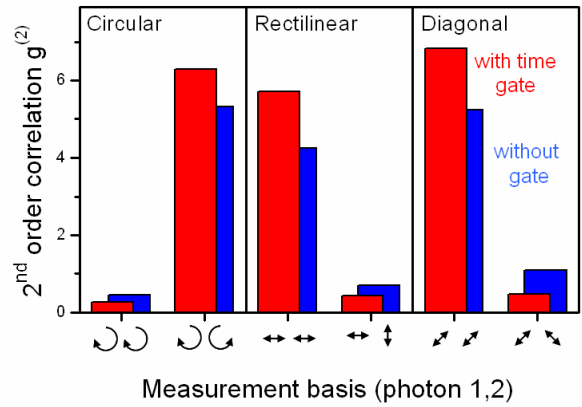

Figure 3. Zero-delay co- and cross- polarised second order correlation measurements between the pair of photons emitted by the biexciton state of a single dot. The red (blue) bars represent data collected with (without) a temporal gate selecting emission from a 1ns window centred on the first photon and a 1.5ns window centred on the second photon.

| Basis | | \|Correlation\| |
|---|---|---|
| ↕ | ⤢ | 0.481 ± 0.050 |
| ⤢ | ⤢ | 0.675 ± 0.048 |
| ↕ | ↔ | 0.611 ± 0.054 |
| ⤢ | ↔ | 0.685 ± 0.041 |
| $S_{CHSH}$ | | **2.45 ± 0.10** |

Table 1: Results of the measurement of the CHSH Bell inequality using the linear polarisation bases as indicated.